\documentclass[twocolumn,aps,nofootinbib,floatfix]{revtex4}

\usepackage{graphicx}
\usepackage{latexsym}
\usepackage{amssymb}
\usepackage{amsmath}
\usepackage{mathrsfs}
\usepackage{wasysym}
\usepackage{textcomp}
\usepackage{braket}
\usepackage{color}
\usepackage[center]{subfigure}

\begin{document}

 \newcommand{\bq}{\begin{equation}}
 \newcommand{\eq}{\end{equation}}
 \newcommand{\bqn}{\begin{eqnarray}}
 \newcommand{\eqn}{\end{eqnarray}}
 \newcommand{\nb}{\nonumber}
 \newcommand{\lb}{\label}
\newcommand{\PRL}{Phys. Rev. Lett.}
\newcommand{\PL}{Phys. Lett.}
\newcommand{\PR}{Phys. Rev.}
\newcommand{\CQG}{Class. Quantum Grav.}
 \newcommand{\sst}{\scriptscriptstyle}
 \newcommand{\lrp}[1]{\left(#1\right)}
 \newcommand{\lrb}[1]{\left[#1\right]}
 \newcommand{\lrc}[1]{\left\{#1\right\}}
 \newcommand{\hong}[1]{\textcolor{red}{#1}}
\title{Primordial Non-Gaussianity of Gravitational Waves in Ho\v{r}ava-Lifshitz Gravity}

\author{Yongqing Huang${} ^{a,b}$}
\email{yongqing_huang@baylor.edu}

\author{Anzhong Wang${} ^{a,b}$}
\email{anzhong_wang@baylor.edu}

\author{Razieh Yousefi${} ^{b}$}
\email{raziyeh_yousefi@baylor.edu}

\author{Tao Zhu${} ^{a,b}$}
\email{tao_zhu@baylor.edu}

\affiliation{ 
${} ^{a}$ Institute for Advanced Physics $\&$ Mathematics, Zhejiang University of
Technology, Hangzhou 310032,  China\\
${} ^{b}$ GCAP-CASPER, Physics Department, Baylor
University, Waco, TX 76798-7316, USA }

\date{\today}

\begin{abstract}
In this paper, we study 3-point correlation function of primordial gravitational waves generated in the de Sitter background in the 
framework of the general covariant Ho\v{r}ava-Lifshitz gravity with an arbitrary coupling constant $\lambda$. We find that, at
cubic order, the interaction Hamiltonian receives contributions from four terms built of the 3-dimensional Ricci tensor $R_{ij}$ 
of the leaves $t = $ constant. In particular, the 3D Ricci scalar $R$ yields the same $k$-dependence as that in general relativity, 
but with different magnitude due to coupling with the $U(1)$ field $A$ and a UV history. Interestingly, the two terms $R_{ij}R^{ij}$ 
and $\left(\nabla^{i}R^{jk}\right)\left(\nabla_{i}R_{jk}\right)$ exhibit peaks at the squeezed limit. We show that this is due to the 
effects of the polarization tensors.  The signal generated by the fourth term, $R^i_j R^j_k R^k_i$, favors the equilateral shape 
when spins of the three tensor fields are the same, but peaks in between the equilateral and squeezed limits when spins are mixed.
The consistency with the recently-released Planck observations on non-Gaussianity is also discussed and is found  that  
$\left(H/M_*\right)^2\left(H/M_{pl}\right) \le 10^{-8}$,  where $M_{*}$ denotes the suppression energy of high-order 
operators,  $M_{pl}$ the Planck mass, and $H$ the energy  of inflation. 
\end{abstract}


\maketitle
\section{Introduction}
\renewcommand{\theequation}{1.\arabic{equation}} \setcounter{equation}{0}

To quantize gravity in the framework of quantum field theory, and take the metric as the fundamental variables, 
recently Ho\v{r}ava proposed the Ho\v{r}ava-Lifshitz (HL) theory of gravity \cite{Horava}.
By construction, the HL theory is power-counting renormalizable, which is realized by including
high-order spatial derivative operators [up to six in (3+1)-dimensional spacetimes].
The exclusion of high-order time derivative operators, on the other hand, ensures
that the theory is unitary, a problem that   high-order  derivative theories of gravity live with for a long time \cite{Stelle}.
Clearly, this inevitably breaks the general  diffeomorphisms,
$\delta{x}^{\mu} = -\zeta^{\mu}(t, x) , (\mu = 0, 1,  2,  3)$.
Although  such a
breaking  in the gravitational sector is much less restricted by experiments/observations than that in the matter sector  \cite{LZbreaking,Pola},
it is still a challenging question how to prevent the propagation of the Lorentz violations into the Standard Model of particle physics \cite{PS}.

Ho\v{r}ava  assumed that such a breaking happens only in the ultraviolet (UV),  and down to the foliation-preserving diffeomorphism, ${\mbox{Diff}}(M, \; {\cal{F}})$,
\bq
\lb{0.2}
 \delta{t} = - f(t), \;\;\; \delta{x}^i = - \zeta^i(t, x), (i = 1, 2, 3).
\eq
 In the infrared (IR), the low derivative operators take over, and the Lorentz symmetry is expected to be ``accidentally" restored, whereby  a
healthy low energy limit is  presumably obtained.

The breaking of the  general  diffeomorphisms immediately results in the appearance of spin-0 gravitons in the theory, in addition to
the spin-2 ones, found in general relativity. This is potentially dangerous, and leads to several problems, including instability,  strong coupling and
different speeds of (massless) particles \cite{reviews}.
To resolve these issues, various models have been proposed, along two 
different lines, one with the projectability condition   \cite{HMT},
$N = N(t)$,
 and the other without it \cite{BPS,ZWWS},  where 
$N$ denotes the lapse function in the Arnowitt, Deser and Misner decompositions  \cite{ADM}. 
In particular,  Ho\v{r}ava and Melby-Thompson (HMT)  proposed to enlarge the foliation-preserving diffeomorphisms  (\ref{0.2})
to include a local U(1)  symmetry, so that the reformulated  theory  has the symmetry  \cite{HMT},
\bq
\lb{symmetry}
 U(1) \ltimes {\mbox{Diff}}(M, \; {\cal{F}}).
 \eq
With such an enlarged symmetry, the spin-0 gravitons, which appear in the original model of the HL theory \cite{Horava}, are eliminated
\cite{HMT, WW}, and as a result, the problems related to them, including instability, ghost, and strong coupling {problems}, are resolved automatically.
This was initially done {in \cite{HMT}} with $\lambda = 1$, where $\lambda$ characterizes the deviation of the
theory from general relativity  in the infrared, as one can see from  Eqs.(\ref{1.1}) and (\ref{1.2}) given below. It was soon generalized
to the case with any $\lambda$ \cite{dS}, in which it was  shown that the spin-0 gravitons are also eliminated   \cite{HW, dS}, so that the above 
mentioned problems are resolved in the gravitational sector even with any $\lambda$.
In the matter sector, the strong coupling problem, first noted  in \cite{HW},  can be solved by introducing a mass $M_{*}$ so that
$M_{*} < \Lambda_{\omega}$, where $M_{*}$ denotes the suppression energy of high order operators, and $\Lambda_{\omega}$
 the would-be energy scale, above which matter becomes strongly coupled
\cite{LWWZ}, similar to the non-projectability case without the enlarged symmetry  \cite{BPSb}. The consistence of this model with solar system 
tests was investigated recently \cite{LMW}, and found that it is consistent with  observations, provided that the gauge field and Newtonian prepotential 
are part of the metric, so that the line element $ds^2$ is invariant not only under  the coordinate transformations 
(\ref{0.2}), but also under the local U(1) gauge transformations,
\bqn
\lb{U(1)}
\delta_\alpha N &=& 0,\;\;\; \delta_\alpha N_i=N\nabla_i\alpha,\;\;\; \delta_\alpha g_{ij}=0,\nb\\
\delta_\alpha A&=& \dot{\alpha}-N^i\nabla_i\alpha,\;\;\;
\delta_\alpha\varphi= - \alpha,
\eqn
where $\alpha$ denotes the U(1) generator, and $\dot{\alpha} \equiv \partial\alpha/\partial t$, and $N^i$ and $g_{ij}$ are, respectively, 
the shift vector  and the 3-metric   of the leaves $t = $ constant, with $N_{i} \equiv g_{ij}N^j$. $A$ denotes the U(1) gauge field, and $\varphi$ the Newtonian  pre-potential.
For detail, see \cite{LMW}. 

A non-trivial generalization of the enlarged symmetry (\ref{symmetry}) to the nonprojectable case $N = N(t, x)$ was recently worked out in \cite{ZWWS},
and showed that the only degree of freedom of the model in the gravitational sector is the spin-2 massless gravitons, the same as that in GR. Because
of the elimination of the spin-0 gravitons, the physically viable region of the coupling constants is considerably enlarged, in comparison with the healthy
extension \cite{BPS}, where   the extra U(1) symmetry is absent.    Furthermore, the
number of independent coupling constants in the gravitational sector  is also dramatically reduced, from about 100 down to 15.  The consistence of the model with cosmology was
showed recently in \cite{ZWWS,ZHW,PPGW,LW}, and various remarkable features were found.

In this paper, we shall work within the HMT framework of the HL theory with the projectability condition $N = N(t)$ \cite{HMT,WW,dS,HW}, although our main
results are expected to hold equally in the non-projectable case \cite{ZWWS}, as the tensor perturbations are almost the same in both cases \cite{HW,ZWWS,ZHW}.

Continuing our previous study of the statistics of the primordial perturbations in single-field slow-roll inflation in the HMT framework \cite{Inflation, NG-HMT}, we study here the 3-point function of the tensor perturbations in de Sitter background. Non-Gaussianity of tensor perturbations have been studied intensively recently in various theories of gravity. In particular, \cite{Gao-PRL} studied the non-Gaussian characteristics in the most general single-field inflation model with second-order field equations and found that the interaction at cubic order is composed of only two terms, which generate squeezed and equilateral shapes. On the other hand, \cite{parity} focused on effects of parity violations. {But, as far as we know, this is the first time to investigate  this problem within the framework of the HL theory.}
 In this paper we shall focus on the shapes of the bispectrum generated by the various terms in the gravity sector, especially the higher spatial derivative terms and leave the topic of parity violations \cite{PPGW} to future studies. For detail on the power spectrum and its scale dependence of the tensor perturbations, we refer readers to section 6.2 of \cite{Inflation}.
 
The rest of the paper is organized as follows. In Section II we give a brief review of the non-relativistic general covariant Ho\v{r}ava-Lifshitz theory of gravity with the projectability condition and an arbitrary coupling constant $\lambda$. {Key results on the power spectrum of tensor perturbations obtained in \cite{Inflation} is also repeated here briefly for convenience.}
The interaction Hamiltonian $H_{\rm I}$ {at cubic order} is analyzed in Section III, and was found to receive contributions from only four terms in the potential part of the theory. 
Section IV performs the integration of the mode functions of the quantized fluctuations, where we also present conditions with which  the UV history in the integral can be ignored as a subleasing error term. We then plot various shapes of the bispectrum generated by the four terms in Section V. In Section VI, we consider the constraints on primordial non-Gaussianity from the recently-release Planck result \cite{planck2013} to obtain  restrictions  on the energy scale $M_*$ and the inflation energy $H$.   Finally, in Section VII we summarize our main results.

\section{General covariant HL gravity with projectability condition}
\renewcommand{\theequation}{2.\arabic{equation}} \setcounter{equation}{0}
The action of the general covariant HL theory of gravity with the projectability condition can be written as \cite{HMT,WW,dS,HW},
\bqn
\lb{1.1}
S &=& \zeta^2 
\int dt d^{3}x N \sqrt{g} \Big({\cal{L}}_{K} -
{\cal{L}}_{{V}} +  {\cal{L}}_{{\varphi}} +  {\cal{L}}_{{A}} +  {\cal{L}}_{{\lambda}}\Big) \nb\\
& & ~~~~ + \int dt d^{3}x N \sqrt{g} {\cal{L}}_{M},
\eqn
where $g={\rm det}(g_{ij})$, $G$ is the Newtonian constant, and
\bqn
\lb{1.2}
{\cal{L}}_{K} &=& K_{ij}K^{ij} -   \lambda K^{2},\\
\lb{1.2v}
{\cal{L}}_{{V}} &=& \zeta^{2}g_{0}  + g_{1} R + \frac{1}{\zeta^{2}}
\left(g_{2}R^{2} +  g_{3}  R_{ij}R^{ij}\right)\nb\\
& & + \frac{1}{\zeta^{4}} \left(g_{4}R^{3} +  g_{5}  R\;
R_{ij}R^{ij}
+   g_{6}  R^{i}_{j} R^{j}_{k} R^{k}_{i} \right)\nb\\
& & + \frac{1}{\zeta^{4}} \left[g_{7}R\Delta R +  g_{8}
\left(\nabla_{i}R_{jk}\right)
\left(\nabla^{i}R^{jk}\right)\right],\\
\lb{lam}
{\cal{L}}_{\varphi} &=&\varphi {\cal{G}}^{ij} \Big(2K_{ij} + \nabla_{i}\nabla_{j}\varphi\Big),\\
{\cal{L}}_{A} &=&\frac{A}{N}\Big(2\Lambda_{g} - R\Big),\\
{\cal{L}}_{\lambda} &=& \big(1-\lambda\big)\Big[\big(\Delta\varphi\big)^{2} + 2 K \Delta\varphi\Big].
\eqn
Here $\lambda$ characterizes the deviation of the theory from general relativity  in the infrared, as mentioned previously,
$\Delta \equiv g^{ij}\nabla_{i}\nabla_{j}$,  $\Lambda_{g}$ is a    coupling constant,  the
Ricci and Riemann tensors $R_{ij}$ and $R^{i}_{jkl}$  all refer to the 3-metric $g_{ij}$, and
\bqn
\lb{1.3}
K_{ij} &=& \frac{1}{2N}\left(- \dot{g}_{ij} + \nabla_{i}N_{j} +
\nabla_{j}N_{i}\right),\nb\\
{\cal{G}}_{ij} &=& R_{ij} - \frac{1}{2}g_{ij}R + \Lambda_{g} g_{ij}.
\eqn
 The   IR limits, on the other hand, 
 require, 
 \bq
 \lb{NC}
 g_{1} = -1,\;\;\; \zeta^2 = \frac{1}{16\pi G} = \frac{1}{2}M_{pl}^2,
 \eq
where 
$M_{pl}$ denotes  the Planck mass.  
  The coupling constants $ g_{s}\, (s=0, 1, 2,\dots 8)$ are all dimensionless, and
 \bq
 \lb{lambda}
 \Lambda = \frac{1}{2}\zeta^2 {g_0},
 \eq
 is the cosmological constant. Since $g_{2,3}$ are the coupling coefficients of the [curvature]$^{2}$ terms, it is expected that
 $g_2/g_2 \simeq {\cal{O}}(1)$. Similarly, it is expected that $g_{4} \simeq g_{5} \simeq ... \simeq g_8$. Then, one can 
 introduce two energy scales,
 \bqn
 \lb{es}
 M_{A} &\equiv& \left(\frac{\hat{g}_{i}}{g_{i}}\right)^{1/2} M_{pl}, \;\; (i = 2, 3), \nb\\
  M_{B} &\equiv& \left(\frac{\hat{g}_{j}}{g_{j}}\right)^{1/4} M_{pl}, \;\; (j = 4, 5, ..., 8),
 \eqn
 where $\hat{g}_{i}, \; \hat{g}_{j} \simeq {\cal{O}}(1)$. In principle, $M_{A}$ and $M_{B}$ are independent, and can be  different \cite{BPS}. However,
 in this paper we  assume that
 \bq
 \lb{m*}
 M_{A} \simeq M_{B} \equiv M_*.
 \eq
${\cal{L}}_{{M}}$ is the Lagrangian of matter fields, and for a scalar field $\chi$, it is given by \cite{WWM,Inflation}, 
\bqn
\lb{1.6a}
{\cal{L}}_{M} &=& {\cal{L}}^{(0)}_{\chi} + {\cal{L}}^{(A,\varphi)}_{\chi},\nb\\
\lb{1.6ab}
{\cal{L}}^{(0)}_{\chi} &=& \frac{f(\lambda)}{2N^{2}}\Big(\dot{\chi} - N^{i}\nabla_{i}\chi\Big)^{2} - {\cal{V}},\\
{\cal V}  &=& V\left(\chi\right) + \left(\frac{1}{2}+V_{1}\left(\chi\right)\right) (\nabla\chi)^2
+  V_{2}\left(\chi\right){\cal{P}}_{1}^{2}\nb\\
& & +  V_{3}\left(\chi\right){\cal{P}}_{1}^{3}  +
V_{4}\left(\chi\right){\cal{P}}_{2} + V_{5}\left(\chi\right)(\nabla\chi)^2{\cal{P}}_{2}\nb\\
& & 
+ V_{6}(\chi) {{\cal P}}_{1} {\cal{P}}_{2}, \\
\lb{1.6b}
{\cal{L}}^{(A,\varphi)}_{\chi} &=& \frac{A - {\cal{A}}}{N}  \Big[c_{1}\left(\chi\right)\Delta\chi + c_{2}\left(\chi\right)\big(\nabla\chi\big)^{2}\Big]\nb\\
&&  - \frac{f}{N}\Big(\dot{\chi} - N^{i}\nabla_{i}\chi\Big)\big(\nabla^{k}\varphi\big)\big( \nabla_{k}\chi\big)\nb\\
& & + \frac{f}{2}\Big[\big(\nabla^{k}\varphi\big)\big(\nabla_{k}\chi\big)\Big]^{2},
 \eqn
 with $c_1\left(\chi\right), c_2\left(\chi\right), V(\chi)$ and $V_{n}(\chi)$ being arbitrary functions of $\chi$, and
\bq
\lb{1.6c}
{\cal{P}}_{n} \equiv \Delta^{n}\chi.
\eq

Note that, similar to the  Lifshitz scalar field \cite{Lifshitz}, now $\chi$ is scaling as $ \chi \rightarrow  \chi$,  under the anisotropic scalings of time and space,
\bq
\lb{scaling}
t  \rightarrow  b^{-3} t,\;\;\; x^i \rightarrow b^{-1} x^i,
\eq
for which the total action $S$ defined by Eq.(\ref{1.1}) is invariant. This also explains why generically $V_{n}(\chi)$ defined above are arbitrary functions of $\chi$.
 
 Applications of the above model to single field slow-roll inflation were studied in \cite{Inflation}, where it was found that the power spectrum of primordial tensor perturbations takes the form
 	\bqn
	 \Delta^2_{\text{T}}(k) &\approx& \Delta^{2\text{GR}}_{\text{T}}(k) \times
	 	\left\{ 
	 	\begin{array}{l l}
		 	\left(1-\frac{9}{2}\epsilon_{{\scriptscriptstyle \text{HL}}}\right), 
				& \quad \text{ $\epsilon_{{\scriptscriptstyle \text{HL}}} \ll 1$,}\\
			\frac{2 e^{1/2} \eta_{{\scriptscriptstyle \text{HL}}}}{9},
				& \quad \text{ $\epsilon_{{\scriptscriptstyle \text{HL}}} \gg 1$,}
		\end{array}
		\right. \nb\\
	r &\simeq& 
	 	\left\{ 
	 	\begin{array}{l l}
		 	16\epsilon_{{\scriptscriptstyle V}}\left(1 - \frac{9}{4}\epsilon_{{\scriptscriptstyle \text{HL}}}\right), 
				& \quad \text{ $\epsilon_{{\scriptscriptstyle \text{HL}}} \ll 1$,}\\
			8\epsilon_{{\scriptscriptstyle V}},
				& \quad \text{ $\epsilon_{{\scriptscriptstyle \text{HL}}} \gg 1$,}
		\end{array}
		\right.
	\eqn
	where $e$ denotes the base of the natural logarithm,     $r \equiv {\Delta^2_{\text{T}}(k)}/{\Delta^2_{{\cal{R}}}(k)}$ is the tensor-to-scalar ratio,  and 
	\bqn
	\Delta^{2\text{GR}}_{\text{T}}(k) &=& \frac{18}{e^3} \frac{2H^2}{\pi^2 M^2_{\text{pl}}}, \nb\\
	\epsilon_{{\scriptscriptstyle \text{HL}}} &\equiv& \frac{H^2}{M^2_*} \equiv 1/ \eta_{{\scriptscriptstyle \text{HL}}}.
	\eqn
The Hubble parameter $H$ takes its value during the slow-roll inflation, so we have $H \equiv H_{{\mbox{inflation}}}$. 
	The overall factor $18/e^3$ in the power spectrum, comparing with the well-known result from the simplest inflation models, is introduced due to the uniform-approximation method that we used. We see that when $\epsilon_{{\scriptscriptstyle \text{HL}}} \ll 1$, the results are consistent with those obtained in the simplest inflation models in general relativity. Hence, for simplicity throughout this paper we   assume  $\epsilon_{{\scriptscriptstyle \text{HL}}} \ll 1$.
 For detail, we refer readers to \cite{HW, Inflation}. 

\section{The Interaction Hamiltonian} 
\renewcommand{\theequation}{3.\arabic{equation}} \setcounter{equation}{0}

Comparing with the two-point correlation function, aka the power spectrum, the non-Gaussian structure of primordial fluctuations, often evaluated with higher-order correlation functions, could provide further important information. In particular, perturbations to both metric and matter need to be performed beyond the linear order in order for the non-Gaussian signature to emerge. Hence,  the scalar, vector and tensor perturbations no longer evolve independently as in the linearized case. And three-point correlation functions, or its Fourier image the bispectrum, is composed of $\langle SSS\rangle, \langle SST\rangle, \langle STT\rangle$, and $\langle TTT \rangle$ etc. \cite{in-in}, where $S$ and $T$ refer to scalar and tensor perturbations. However, due to the smallness of the tensor perturbations as indicated by the not-yet-detected power spectrum of primordial tensor perturbations, attention is usually given to $\langle SSS\rangle$. Here,  as in \cite{Gao-PRL, parity}, we study the 3-point function of three gravitons $\langle TTT\rangle$, in order to obtain  upper bounds on the tensor perturbations, whereby  further constraints on the parameter space of the theory are given, as to be shown in the following sections.

To our goals, we simply assume that no tensor perturbations exist in the matter sector, and that the tensor perturbations of the metric around a spatially flat FLRW metric is \cite{Gao-PRL}
\bq\lb{5t.1.1}
N = a, ~~~ N^i = 0, ~~~ g_{ij} = a^2 \lrp{e^h}_{ij},
\eq
where
\bq\lb{5t.1.1a}
\lrp{e^h}_{ij} = \delta_{ij} + h_{ij} + \frac{1}{2} h_{ik}h^{k}_j + \frac{1}{6} h_{ij}h^{jk}h_{ki} + \cdots.
\eq
The metric perturbation is defined in this way such that there is no cubic term involving two time derivatives in general relativity  \cite{in-in}. The small dimensionless quantity $h_{ij}$ satisfies the transverse and traceless condition $\partial^i h_{ij} = 0 = h^i_i$. Moreover, its indices are lowered and raised by $\delta_{ij}$ and $\delta^{ij}$. Thus for simplicity of notations, in this paper we shall not distinguish super-indices and sub-indices, with the understanding   that when an index appears twice, summation over that index is performed. We further introduce a short hand notation
\bq\lb{5t.1.1b}
\lrp{h^n}_{ij} \equiv h_{ik_{\sst{1}}}h_{k_{\sst{1}}k_{\sst{2}}} \cdots h_{k_{\sst{n-1}}j}.
\eq

With these perturbations, in the cubic order ${\cal{O}}\left(h^3\right)$ the interaction Hamiltonian  is found to receive contributions only from four terms,
\bq\lb{5t.2.1}
R, ~~~ R_{ij}R^{ij}, ~~~ R^i_j R^j_k R^k_i, ~~~ \left(\nabla^{i}R^{jk}\right)\left(\nabla_{i}R_{jk}\right).
\eq
In this order, the kinetic part of the action $S_K$, like in the case of general relativity, does not contribute to $H_I$ even though $\lambda$ now could differ from 1 [cf. Eq.(\ref{1.2})]. We find that,
\bqn\lb{5t.2.2}
&&  R  = \frac{a^{-2}}{4} h^{ij,ab}\left( 2h_{ia}h_{jb} -  h_{ij}h_{ab}\right), \\[2mm]
&&  R_{ij}R^{ij} = 
a^{-4} \left(\partial^2h^{ij}\right)\Bigg\{\Big[\frac{1}{2}h_{ia,b}h_{jb,a}\Big] \nb\\
 && ~~~~~~~~~~~ +  h^{ab}\Big[h_{ia,jb}-\frac{1}{4}h_{ab,ij}   -\frac{1}{2}h_{ij,ab}\Big]\Bigg\}, \\[2mm]
&& R^i_j R^j_k R^k_i  = -\frac{a^{-6}}{8} \left[\partial^2 \left(h _{ij}\right)\right]\left[\partial^2 \left(h _{jk}\right)\right]\left[\partial^2 \left(h _{ki}\right)\right],
~~~~~~ \\[2mm]
&& (\nabla^{i}R^{jk})(\nabla_{i}R_{jk})  =  \frac{a^{-6}}{4} \left(\partial^2h_{ij}\right)\Big[h^{ab}\left(\partial^2h_{ij}\right)_{,ab}\nb\\
&& ~~~~~~~~~~~~~~~~~~~~~~~~~~ - 4 h^{ai,b}\left(\partial^2h_{bj}\right)_{,a}\Big]\nb\\
 &&  ~~~~~~~ - \frac{a^{-6}}{4} \left(\partial^4h_{ij}\right) \Bigg\{2\Big[h_{ia,b}h_{jb,a}\Big]\nb\\
 && ~~~~~~~ -  h^{ab}\Big[-4h_{ia,bj}+2h_{ij,ab}+h_{ab,ij}\Big]\Bigg\}.
\eqn

The contribution from $R$ is the same as that obtained in \cite{Gao-PRL}. The contribution from $R^i_j R^j_k R^k_i$ differs from theirs in an essential way. The model considered in \cite{Gao-PRL} describes the class of single-field inflation models where the coupling between the inflaton and gravity could be different from the canonical form while Lorentz symmetry is kept. This explains why their interaction term posses time-derivatives. On the other hand, coupling of the gravitational sector with scalar matter in our model is in the canonical form [cf. Eq. (\ref{1.6ab})]. The higher spatial derivative terms exist because of the Lorentz symmetry breaking.

We define the Fourier image of $h_{ij}$ here in the canonical  form,
\bq
h_{ij} = \int \frac{d^3 k}{(2 \pi)^3} \sum_{s=+, -} \varepsilon^s_{ij}(\mathbf{k}) h^s_\mathbf{k}(t) e^{i\mathbf{k}\mathbf{x}},
\eq
where $h^s_\mathbf{k}(t)$ is now a scalar quantity, and the rank-2 polarization tensors satisfy $\varepsilon^s_{ii}(\mathbf{k}) = k^i \varepsilon^s_{ij} (\mathbf{k})=0$ and $\varepsilon^s_{ij}(\mathbf{k}) \varepsilon^{s'}_{ij}(\mathbf{k})=\delta_{ss'}$.\footnote{This set of polarization tensors is related to those introduced in \cite{Inflation} through
\bq\lb{5t.2.3}
\varepsilon^\pm_{ij} = \frac{1}{\sqrt{2}} \lrp{\epsilon^+_{ij} \pm i \epsilon^\times_{ij}}.
\eq}
By a proper choice of phase, they satisfy the relation \cite{Gao-PRL}
\bq\lb{5t.2.3a}
\lrb{\varepsilon^{s}_{ij}(\mathbf{k})}^* = \varepsilon^{-s}_{ij}(\mathbf{k}) = \varepsilon^{s}_{ij}(-\mathbf{k}).
\eq
Then, in the momentum space we find
\bqn
\lb{5t.2.5a}
&&a^2R \leadsto -\frac{1}{6}\left\{\Big[C_1\Big]_{\mathbf{q}}(123) - \frac{1}{4} \Big[C_2\Big]_{\mathbf{q}}(123) \right.\nb\\
					&& ~~~~~~~~~~~~~~~~~~~~~\left. - \frac{1}{4} \Big[C_3\Big]_{\mathbf{q}}(123)\right\} + {\rm cyclic},\\
\lb{5t.2.5b}
&&a^4 R^{ij}R_{ij}\leadsto \frac{q^2_1 + q^2_2 + q^2_3}{6} \left\{\Big[C_1\Big]_{\mathbf{q}}(123) \right. \nb\\
					&& ~~~~~~~~~~~~~~~~~\left. - \frac{1}{4} \Big[C_2\Big]_{\mathbf{q}}(123)  - \frac{1}{4} \Big[C_3\Big]_{\mathbf{q}}(123)\right\} \nb\\
					&&   ~~~~~~~~~~~~~~~~~ + {\rm cyclic},\\
\lb{5t.2.5c}
&&a^6 R^i_j R^j_k R^k_i \leadsto \frac{\lrp{q_1 q_2 q_3}^2}{24} \Big[C_5\Big]_{\mathbf{q}}(123)\nb\\
&&  ~~~~~~~~~~~~~~~~~~~~  + {\rm cyclic},\\
\lb{5t.2.5d}
&& a^6\left(\nabla^{i}R^{jk}\right)\left(\nabla_{i}R_{jk}\right) 
				\leadsto \frac{q^4_1 + q^4_2 + q^4_3}{6} \left\{\Big[C_1\Big]_{\mathbf{q}}(123)  \right.\nb\\
					&& ~~~~~~  \left. - \frac{1}{4} \Big[C_2\Big]_{\mathbf{q}}(123) - \frac{1}{4} \Big[C_3\Big]_{\mathbf{q}}(123)\right\} \nb\\
					&& ~~~~~~ + \frac{q^2_1q^2_3}{6} \left\{\Big[C_1\Big]_{\mathbf{q}}(123) - \frac{1}{4} \Big[C_3\Big]_{\mathbf{q}}(123)\right\}\nb\\
					&& ~~~~~~ + \frac{q^2_1q^2_2}{6} \left\{\Big[C_1\Big]_{\mathbf{q}}(123) - \frac{1}{4} \Big[C_2\Big]_{\mathbf{q}}(123)\right\}\nb\\
					&& ~~~~~~  + {\rm cyclic},
\eqn
where ``cyclic" refers to the cyclic rotation of (1, 2, 3), and  we have defined the symbol $\leadsto$ to have meaning of
\bq\lb{5t.2.6}
= \int \prod^3_{j=1} \lrb{\frac{d^3 q_{\scriptscriptstyle j}e^{i \mathbf{x} \cdot \mathbf{q}_{\scriptscriptstyle j}}}{\left(2\pi\right)^{3}}\sum_{s_{\scriptscriptstyle j}=+,-} h^{s_{\scriptscriptstyle j}}_{\mathbf{q}_{\scriptscriptstyle j}}\left(t'\right)},
\eq
and introduced the shorthand notations
\bqn\lb{5t.2.8}
\Big[C_1\Big]_{\mathbf{q}}(123) &\equiv& k_{\sst{1a}} k_{\sst{1b}}
\varepsilon^{s_{\scriptscriptstyle 1}}_{ij}\left(\mathbf{q}_{\scriptscriptstyle 1}\right)\varepsilon^{s_{\scriptscriptstyle 2}}_{ia}\left(\mathbf{q}_{\scriptscriptstyle 2}\right)\varepsilon^{s_{\scriptscriptstyle 3}}_{jb}\left(\mathbf{q}_{\scriptscriptstyle 3}\right),\nb\\
\Big[C_2\Big]_{\mathbf{q}}(123) &\equiv& k_{\sst{1a}} k_{\sst{1b}}
\varepsilon^{s_{\scriptscriptstyle 1}}_{ij}\left(\mathbf{q}_{\scriptscriptstyle 1}\right)\varepsilon^{s_{\scriptscriptstyle 2}}_{ab}\left(\mathbf{q}_{\scriptscriptstyle 2}\right)\varepsilon^{s_{\scriptscriptstyle 3}}_{ij}\left(\mathbf{q}_{\scriptscriptstyle 3}\right),\nb\\
\Big[C_3\Big]_{\mathbf{q}}(123) &\equiv& k_{\sst{1a}} k_{\sst{1b}}
\varepsilon^{s_{\scriptscriptstyle 1}}_{ij}\left(\mathbf{q}_{\scriptscriptstyle 1}\right)\varepsilon^{s_{\scriptscriptstyle 2}}_{ij}\left(\mathbf{q}_{\scriptscriptstyle 2}\right)\varepsilon^{s_{\scriptscriptstyle 3}}_{ab}\left(\mathbf{q}_{\scriptscriptstyle 3}\right),\nb\\
\Big[C_4\Big]_{\mathbf{q}}(123) &\equiv& k_{\sst{1a}} k_{\sst{1b}}
\varepsilon^{s_{\scriptscriptstyle 1}}_{ij}\left(\mathbf{q}_{\scriptscriptstyle 1}\right)\varepsilon^{s_{\scriptscriptstyle 2}}_{ab}\left(\mathbf{q}_{\scriptscriptstyle 2}\right)\varepsilon^{s_{\scriptscriptstyle 3}}_{ab}\left(\mathbf{q}_{\scriptscriptstyle 3}\right),\nb\\
\Big[C_5\Big]_{\mathbf{q}}(123) &\equiv&
\varepsilon^{s_{\scriptscriptstyle 1}}_{ij}\left(\mathbf{q}_{\scriptscriptstyle 1}\right)\varepsilon^{s_{\scriptscriptstyle 2}}_{jk}\left(\mathbf{q}_{\scriptscriptstyle 2}\right)\varepsilon^{s_{\scriptscriptstyle 3}}_{ki}\left(\mathbf{q}_{\scriptscriptstyle 3}\right).
\eqn

Hence, the Hamiltonian reads
\bqn\lb{5t.2.7}
H_{\rm I}(t') &=& \frac{(2\pi)^3\delta(\mathbf{q}_{\scriptscriptstyle 1}+\mathbf{q}_{\scriptscriptstyle 2}+\mathbf{q}_{\scriptscriptstyle 3})}{a^{-3}\zeta^2}\int \prod^3_{j=1} \frac{d^3 q_{\scriptscriptstyle j}}{\left(2\pi\right)^{3}} 
		\sum_{s_{\scriptscriptstyle j}} h^{s_{\scriptscriptstyle j}}_{\mathbf{q}_{\scriptscriptstyle j}}\left(t'\right)
		\nb\\
		&& \times\Bigg[\left(\bar{A} - 1\right)R + \frac{g_3}{\zeta^2}\left(R^{ij}R_{ij}\right) \nb\\
		&& ~~~~ + \frac{g_8}{\zeta^4}\left(\nabla^{i}R^{jk}\right)\left(\nabla_{i}R_{jk}\right) + \frac{g_6}{\zeta^4}R^i_j R^j_k R^k_i\Bigg], \nb\\
\eqn
where expressions of $R, \left(R^{ij}R_{ij}\right), (\nabla^{i}R^{jk})(\nabla_{i}R_{jk})$ and $R^i_j R^j_k R^k_i$ are given in Eqs.(\ref{5t.2.5a}-\ref{5t.2.5d}).

Promoting the scalar variable $h^s_\mathbf{k}(t)$ to a quantized field
\bq\lb{5t.3.1}
\hat{h}^s_{\mathbf{k}}(t) = h_{\mathbf{k}}(t)\hat{a}_s(\mathbf{k}) + h^*_{-\mathbf{k}}(t)\hat{a}^\dagger_s(-\mathbf{k}),
\eq
where the creation and annihilation operators satisfy the commutation relation
\bq
\lrb{\hat{a}_s(\mathbf{k}), \hat{a}^\dagger_{s'}(\mathbf{k}')} = \lrp{2\pi}^3 \delta(\mathbf{k}-\mathbf{k}')\delta_{ss'},
\eq
we are now in the position to employ the in-in formalism \cite{in-in}.  In particular, the 3-point correlator we seek for is given by
\bqn\lb{5t.3.2}
&&\langle \hat{h}^{s_{\scriptscriptstyle 1}}_{\mathbf{k}_{\scriptscriptstyle 1}}\left(t\right) \hat{h}^{s_{\scriptscriptstyle 2}}_{\mathbf{k}_{\scriptscriptstyle 2}}\left(t\right) \hat{h}^{s_{\scriptscriptstyle 3}}_{\mathbf{k}_{\scriptscriptstyle 3}}\left(t\right) \rangle \nb\\
					&\simeq& 
i\int^t_{t_i} dt' \langle\left[\hat{H}_{\rm I}(t'), \hat{h}^{s_{\scriptscriptstyle 1}}_{\mathbf{k}_{\scriptscriptstyle 1}}\left(t\right) \hat{h}^{s_{\scriptscriptstyle 2}}_{\mathbf{k}_{\scriptscriptstyle 2}}\left(t\right) \hat{h}^{s_{\scriptscriptstyle 3}}_{\mathbf{k}_{\scriptscriptstyle 3}}\left(t\right)\right]\rangle \nb\\
					&=&
i \lrp{2\pi}^3 \delta \lrp{\mathbf{k}_{\sst{1}} + \mathbf{k}_{\sst{2}} + \mathbf{k}_{\sst{3}}} \int^t_{t_i} d t' \frac{a^3(t')}{\zeta^2} \nb\\
					&& \times
\Big\{F_{s_{\sst 1}s_{\sst 2}s_{\sst 3}} \lrp{-\mathbf{K}, t'} \Big[ W\lrp{\mathbf{K}, t'; t} - W^*\Big]\Big\}\nb\\
&&~~~~~ + \text{ 5 permutations of } (\mathbf{k}_1, \mathbf{k}_2, \mathbf{k}_3),
\eqn
where we have defined the product of mode functions
\bq\lb{5t.3.4}
W\lrp{\mathbf{K}, t'; t} \equiv h_{k_{\sst{1}}}(t') h^*_{k_{\sst{1}}}(t) h_{k_{\sst{2}}}(t') h^*_{k_{\sst{2}}}(t) h_{k_{\sst{3}}}(t')h^*_{k_{\sst{3}}}(t),
\eq
and for the contraction $\{(\mathbf{q}_1, \mathbf{k}_1), (\mathbf{q}_2, \mathbf{k}_2), (\mathbf{q}_3, \mathbf{k}_3)\}$
\bqn\lb{5t.3.3}
&& F_{s_{\sst 1}s_{\sst 2}s_{\sst 3}} \lrp{-\mathbf{K}, t'} \nb\\
	&\equiv& \lrp{\frac{\bar{A} - 1}{a^2(t')}} \frac{1}{6} \Big[C_1 - \frac{1}{4} C_2 - \frac{1}{4} C_3\Big]^*_{\mathbf{k}}(123)\nb\\
	&& - \frac{g_3}{a^4(t')\zeta^2} \frac{k^2_1 + k^2_2 + k^2_3}{6} \Big[C_1 - \frac{1}{4} C_2 - \frac{1}{4} C_3\Big]^*_{\mathbf{k}}(123)\nb\\
	&& - \frac{g_8}{a^6(t')\zeta^4} \frac{k^4_1 + k^4_2 + k^4_3}{6} \Big[C_1 - \frac{1}{4} C_2 - \frac{1}{4} C_3\Big]^*_{\mathbf{k}}(123)\nb\\
	&& -  \frac{g_8}{a^6(t')\zeta^4} \frac{k^2_1 k^2_3}{6} \Big[C_1 - \frac{1}{4} C_2\Big]^*_{\mathbf{k}}(123) \nb\\
	&& -  \frac{g_8}{a^6(t')\zeta^4} \frac{k^2_1 k^2_2}{6} \Big[C_1 - \frac{1}{4} C_3\Big]^*_{\mathbf{k}}(123) \nb\\
	&& + \frac{g_6}{a^6(t')\zeta^4} \frac{k^2_1  k^2_2  k^2_3}{24} \Big[C_5\Big]^*_{\mathbf{k}}(123) + \text{cyclic},
\eqn
with 
\bqn
\lb{5t.3.3a}
\Big[C_1\Big]^*_{\mathbf{k}}(123) &\equiv& k_{\sst{1a}} k_{\sst{1b}}
\varepsilon^{s_{\scriptscriptstyle 1}}_{ij}\left(-\mathbf{k}_{\scriptscriptstyle 1}\right)\varepsilon^{s_{\scriptscriptstyle 2}}_{ia}\left(-\mathbf{k}_{\scriptscriptstyle 2}\right)\varepsilon^{s_{\scriptscriptstyle 3}}_{jb}\left(-\mathbf{k}_{\scriptscriptstyle 3}\right)\nb\\
&=& k_{\sst{1a}} k_{\sst{1b}}
\varepsilon^{-s_{\scriptscriptstyle 1}}_{ij}\left(\mathbf{k}_{\scriptscriptstyle 1}\right)\varepsilon^{-s_{\scriptscriptstyle 2}}_{ia}\left(\mathbf{k}_{\scriptscriptstyle 2}\right)\varepsilon^{-s_{\scriptscriptstyle 3}}_{jb}\left(\mathbf{k}_{\scriptscriptstyle 3}\right)\nb\\
&=& k_{\sst{1a}} k_{\sst{1b}}
\Big[\varepsilon^{s_{\scriptscriptstyle 1}}_{ij}\left(\mathbf{k}_{\scriptscriptstyle 1}\right)\varepsilon^{s_{\scriptscriptstyle 2}}_{ia}\left(\mathbf{k}_{\scriptscriptstyle 2}\right)\varepsilon^{s_{\scriptscriptstyle 3}}_{jb}\left(\mathbf{k}_{\scriptscriptstyle 3}\right)\Big]^*, ~~~~~~
\eqn
and similarly for $\Big[C_2\Big]^*_{\mathbf{k}}(123)$ and $\Big[C_3\Big]^*_{\mathbf{k}}(123)$. We see that the first line of (\ref{5t.3.3}) is contributed by $R$, the second line by $R^{ij}R_{ij}$, the third, fourth and fifth lines by $\left(\nabla^{i}R^{jk}\right)\left(\nabla_{i}R_{jk}\right)$, and the last line by $R^i_j R^j_k R^k_i$.

\section{The Mode Integration}
\renewcommand{\theequation}{4.\arabic{equation}} \setcounter{equation}{0}

As was noted in \cite{NG-HMT}, the $k$-dependence of the bispectrum receives contributions from both the interaction Hamiltonian and the mode function, whose behavior is determined by the EoM  and its initial state (mathematically the initial condition). In principle, one should perform the integration in (\ref{5t.3.2}) tracing the behavior of the mode function through out its history from the initial time ($t_i$) until several e-folds after horizon exit when the mode freezes ($t$). However, during the early times the modes are highly oscillatory and the integration during that time should be mostly canceled out by itself. This, of course, requires some conditions on the integrands. 

Looking at the integrands in (\ref{5t.3.2}), we see that the non-oscillating time-dependent parts come from the common factor $a^3(t')$, the factor of $a^{-2n}(t')$ in $F_{s_{\sst 1}s_{\sst 2}s_{\sst 3}}$ ($n = 1, 2, 3$)  and from the three mode functions. The oscillating time-dependent part in the mode function is modeled as \cite{NG-HMT}
\bq
h \propto (\eta')^m \exp[i \lambda_l k^l (\eta')^l],
\eq
where $m$ and $l$ are dynamical, i.e. they change when different terms dominate the dispersion relation. 
Since we are working with de Sitter space, $a = -1/(H\eta)$, hence the integration can be written in a schematic form
\bq
\int d\eta' \exp[i \lambda_l \lrp{k^l_1 + k^l_2 + k^l_3} \eta'^l] \eta'^{3m + (2n - 4)}.
\eq
When $3m + (2n - 4) \le 0$, the corresponding integration must be small because  $\eta'^{3m + (2n - 4)}$ changes very slowly during the early times ($\eta' \to -\infty$) while the oscillating part has high frequencies; similarly when $0 < 3m + (2n - 4) < l$, the oscillation smoothes out the changes in $\eta'^{3m + (2n - 4)}$ as one can always perform a change of variables $\tau = \eta'^{3m + (2n - 4) + 1}$ and integrate over $\tau$ with a purely oscillating integrand. When $3m + (2n - 4) \ge l$, on the other hand, care is needed. However, if $3m + (2n - 4)$ is not so higher than $l$, the errors introduced by ignoring this history should not be too large. In fact for the current model, when $k^6$ term dominates the dispersion relation, $m=0, l = 3$ and $n_{\sst{\rm MAX}} = 3$, hence $3m + (2n - 4) \le 2 < l$. Therefore we shall work under this assumption below, and consider only the period when the $k^2$ term dominates the dispersion \footnote{It should be noted that this assumption also implies that   
$H \ll M_*$, so that  quantum effects  are negligible. Recall that $H$ denotes the  energy of inflation.}. During this period, the oscillating mode function  takes the form, 
\bqn
\lb{modes-r}
v_k (t) &=& \frac{C_{+}(k\eta_i)}{\sqrt{2k}} \lrp{1 - \frac{1}{i c_{\sst{T}}k\eta}} e^{-ic_{\sst{T}}k\eta} \nb\\
			&& + \frac{C_{-}(k\eta_i)}{\sqrt{2k}} \lrp{1 + \frac{1}{ic_{\sst{T}}k\eta}} e^{ic_{\sst{T}}k\eta},
\eqn
where $c^2_{\sst{T}} \equiv (1- \bar{A})$, the canonically normalized field $v_k(t)$ is related to $h_k(t)$ through\footnote{Note that the relation here is different from the relation in \cite{Inflation} due to the different normalization we choose for the polarization tensors. The EoM's however, are the same for both.}
\bq
v_k(t) = \frac{a M_{\rm pl}}{2} h_k(t),
\eq
and the constants $C_+$ and $C_{-}$ are in general functions of $H$, $M_{\rm pl}$, $M_{*}$ and transition times $\eta_{\sst{\rm tr}}$ \cite{NG-HMT}. 
The reason for this dependence on $(k\eta_{\sst{\rm tr}})$ is the existence of the UV stage when the dispersion relation differs from the relativistic form significantly. The detailed form however, depends on the procedure of matching of this ``relativistic solution" with the solution in the UV region. One example is the matching we considered in \cite{NG-HMT}.

Summing up the discussions, we see that the UV history has at least three effects on the integration we are considering: one is that $\eta_i$ can no longer be extended to Euclidean space $-\infty(1 + i\epsilon)$; second, the dependence of the normalization on $\eta_{\sst{\rm tr}}$; third, if the mode underwent a non-adiabatic period ($\omega^2 < 2/\eta^2$), then a ``negative frequency" branch exists ($C_{-} \neq 0$). Since  we have seen in \cite{NG-HMT} that a mixture of ``negative frequency" and ``positive frequency" modes gives an enhanced folded shape in the bispectrum \cite{NG-HMT}, in this paper we focus on the case when $C_{-} = 0$.

The bispectrum is then given as
\bqn
\lb{bs}
&&\langle \hat{h}^{s_{\scriptscriptstyle 1}}_{\mathbf{k}_{\scriptscriptstyle 1}}\left(t\right) \hat{h}^{s_{\scriptscriptstyle 2}}_{\mathbf{k}_{\scriptscriptstyle 2}}\left(t\right) \hat{h}^{s_{\scriptscriptstyle 3}}_{\mathbf{k}_{\scriptscriptstyle 3}}\left(t\right) \rangle \nb\\
					&=&
 \lrp{2\pi}^3 \delta \lrp{\mathbf{k}_{\sst{1}} + \mathbf{k}_{\sst{2}} + \mathbf{k}_{\sst{3}}} {\zeta^{-2}} \left|\frac{\sqrt{2}HC_{+}}{c_{\sst{T}}M_{\sst{\rm pl}}}\right|^6 \frac{2G^*_{s_{\sst 1}s_{\sst 2}s_{\sst 3}} \lrp{\mathbf{K}}}{k^3_1k^3_2k^3_3}
\nb\\&&
+ \text{5 perm.'s} + {\cal{E}}({\rm UV}) + {\cal{E}}({\rm Finite} ~\eta_i),
\eqn
where we have introduced two error terms emphasizing that we considered only the relativistic region in the integration, and defined, for the contraction pairing $\{(\mathbf{q}_1, \mathbf{k}_1), (\mathbf{q}_2, \mathbf{k}_2), (\mathbf{q}_3, \mathbf{k}_3)\}$,
\bqn
\lb{5t.3.G}
&&~~ G^*_{s_{\sst 1}s_{\sst 2}s_{\sst 3}} \lrp{\mathbf{K}} \nb\\ &&\equiv \lrp{\bar{A} - 1} {\cal{I}}  \frac{1}{6} \Big[C_1 - \frac{1}{4} C_2 - \frac{1}{4} C_3\Big]^*_{\mathbf{k}}(123)\nb\\
	&&~~ - \frac{g_3}{\zeta^2}{\cal{II}} \frac{k^2_1 + k^2_2 + k^2_3}{6} \Big[C_1 - \frac{1}{4} C_2 - \frac{1}{4} C_3\Big]^*_{\mathbf{k}}(123)\nb\\
	&&~~ - \frac{g_8}{\zeta^4}{\cal{III}} \frac{k^4_1 + k^4_2 + k^4_3}{6} \Big[C_1 - \frac{1}{4} C_2 - \frac{1}{4} C_3\Big]^*_{\mathbf{k}}(123)\nb\\
	&&~~ -  \frac{g_8}{\zeta^4}{\cal{III}} \frac{k^2_1 k^2_3}{6} \Big[C_1 - \frac{1}{4} C_2\Big]^*_{\mathbf{k}}(123) \nb\\
	&&~~ -  \frac{g_8}{\zeta^4}{\cal{III}} \frac{k^2_1 k^2_2}{6} \Big[C_1 - \frac{1}{4} C_3\Big]^*_{\mathbf{k}}(123) \nb\\
	&&~~ + \frac{g_6}{\zeta^4}{\cal{III}} \frac{k^2_1  k^2_2  k^2_3}{24} \Big[C_5\Big]^*_{\mathbf{k}}(123) + \text{cyclic},
\eqn
where
\bqn
\lb{intte}
{\cal{I}} &=& \frac{- \lrp{k_1 + k_2 + k_3}^3 + k_1k_2k_3}{c^{-1}_{\sst{T}} \lrp{k_1 + k_2 + k_3}^2}\nb\\
			&& + \frac{\lrp{k_1 + k_2 + k_3}\lrp{k_1k_2 + k_2k_3 + k_3k_1}}{c^{-1}_{\sst{T}} \lrp{k_1 + k_2 + k_3}^2},\nb\\
{\cal{II}} &=& -2 \frac{\lrp{k_1 + k_2 + k_3}^3 + 3k_1k_2k_3}{c_{\sst{T}}\lrp{k_1 + k_2 + k_3}^4}\nb\\
		&&	- 2\frac{\lrp{k_1 + k_2 + k_3}\lrp{k_1k_2 + k_2k_3 + k_3k_1}}{c_{\sst{T}}\lrp{k_1 + k_2 + k_3}^4},\nb\\
{\cal{III}} &=& 8 \frac{\lrp{k_1 + k_2 + k_3}^3 + 15 k_1k_2k_3}{c_{\sst{T}}^3\lrp{k_1 + k_2 + k_3}^6},\nb\\
		&& + 24 \frac{\lrp{k_1 + k_2 + k_3}\lrp{k_1k_2 + k_2k_3 + k_3k_1}}{c_{\sst{T}}^3\lrp{k_1 + k_2 + k_3}^6}. ~~~~
\eqn

We see that the magnitude of the bispectrum depends on $C_{+}$ which in turn depends on the  new energy scale $M_*$. Since  we have found in \cite{Inflation} that the general relativistic value of the power spectrum for tensor perturbations is obtainable in the HMT framework when $H \ll M_*$, we can have the same conclusion as that for scalar perturbations presented in \cite{NG-HMT}: a large bispectrum is possible, provided that $M_*$ is not too much lower than $M_{\rm pl}$.
\section{Shape of the Bispectrum}
\renewcommand{\theequation}{5.\arabic{equation}} \setcounter{equation}{0}

We are now ready to plot the shapes of the bispectrum. For $s_1 = s_2 = s_3 = 1$ and $s_1 = s_2 = -s_3 = 1$, we plot the shapes contributed by  various terms separately in Figs. \ref{fig5t+} and \ref{fig5t-}. These two figures represent all possible configurations when we do not have parity violating terms. 

When spins of all 3 tensor fields are the same ($s_1 = s_2 =  s_3  = 1$), the signal generated by the cubic terms ${\cal{O}}\left(h^3\right)$ of  $R$ peaks at the squeezed limit ($k_3/k_1 \to 0$),  while 
the ones from  $R^i_j R^j_k R^k_i$ favors the equilateral shape ($k_2/k_1 \simeq k_3/k_1 \simeq 1$). 

It's interesting to note that the other two terms, $R^{ij}R_{ij}$ and $\left(\nabla^{i}R^{jk}\right)\left(\nabla_{i}R_{jk}\right)$, also generate larger signal in the squeezed limit, though they are of higher-order derivatives. {This is indeed expected, if one realizes that the $k$-dependence of them are similar to that of the general relativity  term [cf. Eqs.(\ref{5t.2.5a}-\ref{5t.2.5d})].}
To illustrate this, we plot the $k$-dependence of them in both cases ($+++$) and ($++-$)  in Figs. \ref{fig5t+.basis} and \ref{fig5t-.basis}, where  we have defined
\bqn\lb{confu}
{\rm Configuration 1} &=& {\Big[C_1 - \frac{1}{4} C_2 - \frac{1}{4} C_3\Big]^*_{\mathbf{k}}(123)} + ~ {\rm cyclic}, \nb\\
{\rm Configuration 2} &=& k^2_1 k^2_3 {\Big[C_1 - \frac{1}{4} C_2\Big]^*_{\mathbf{k}}(123)} + ~ {\rm cyclic}, \nb\\
{\rm Configuration 3} &=& k^2_1 k^2_2 {\Big[C_1 - \frac{1}{4} C_3\Big]^*_{\mathbf{k}}(123)} + ~ {\rm cyclic}, \nb\\
{\rm Configuration 4} &=& \Big[C_5\Big]^*_{\mathbf{k}}(123) + ~ {\rm cyclic}.
\eqn

Looking at Eqs. (\ref{bs} - \ref{intte}) and Fig. \ref{fig5t+.basis}, one could already see that for the two terms $R$ and $R_{ij} R^{ij}$, the signal would peak at the squeezed limit. The reason is the following. The first line of (\ref{5t.3.G}) is contributed by $R$. The $k$-dependence of the GR effect comes from the configuration 1, the function ${\cal{I}}$ resulted from integration, and the common factor $(k_1k_2k_3/k^3_1)^{-1}$ in plotting Fig. \ref{fig5t+}.\footnote{This factor is necessary to make the quantity we plot as dimensionless. The actual factor is $k^3_1(k_1k_2k_3/k^3_1)^{-1}$. However, as the figure is normalized w.r.t. the equilateral configuration, $k^3_1$ can be safely dropped.} We see from panel (a) of Fig. \ref{fig5t+.basis} that configuration 1 has a higher magnitude at the equilateral omit relative to the squeezed limit--detailed calculations show that the ratio is approximately 4:1. This ratio is magnified by a factor of 1.3 considering the function ${\cal{I}}$. However, the common factor $(k_1k_2k_3/k^3_1)^{-1}$ makes it clear that, the overall shape contributed by the GR term must be of the squeezed. For the contributions of the term  $R_{ij} R^{ij}$, the analysis is similar, which is represented by the second line in (\ref{5t.3.G}). The same is also true for $\left(\nabla^{i}R^{jk}\right)\left(\nabla_{i}R_{jk}\right)$, which is responsible for generating the third, fourth and fifth lines in (\ref{5t.3.G}).

The case for $R^i_j R^j_k R^k_i$--the last line in (\ref{5t.3.G})--is different. The effect of the common factor $(k_1k_2k_3/k^3_1)^{-1}$ is cancelled out because of the factor $k^2_1k^2_2k^2_3$ in (\ref{5t.3.G}). Hence the overall shape, after taking into account of the shape of configuration 4 and function ${\cal{III}}$, is of the equilateral.

%

When the spins are mixed ($s_1 = s_2 =  -s_3  = 1$), the terms $R, R^{ij}R_{ij}$ and $\left(\nabla^{i}R^{jk}\right)\left(\nabla_{i}R_{jk}\right)$ generate shapes similar to the previous case. A particularly interesting result is that the signal generated by $R^i_j R^j_k R^k_i$ no longer favors the equilateral shape but peaks in between the equilateral and squeezed limits. This can be understood as that for the mixed spin case, the product of the polarization tensors (in particular, configuration 4 defined above, which appear in the contribution from $R^i_j R^j_k R^k_i$) gives a strong favor of the squeezed shape, as seen in Fig. \ref{fig5t-.basis}.

\begin{figure}
\centering
	\subfigure[]
	{\label{fig:5t+a}\includegraphics[width=55mm]{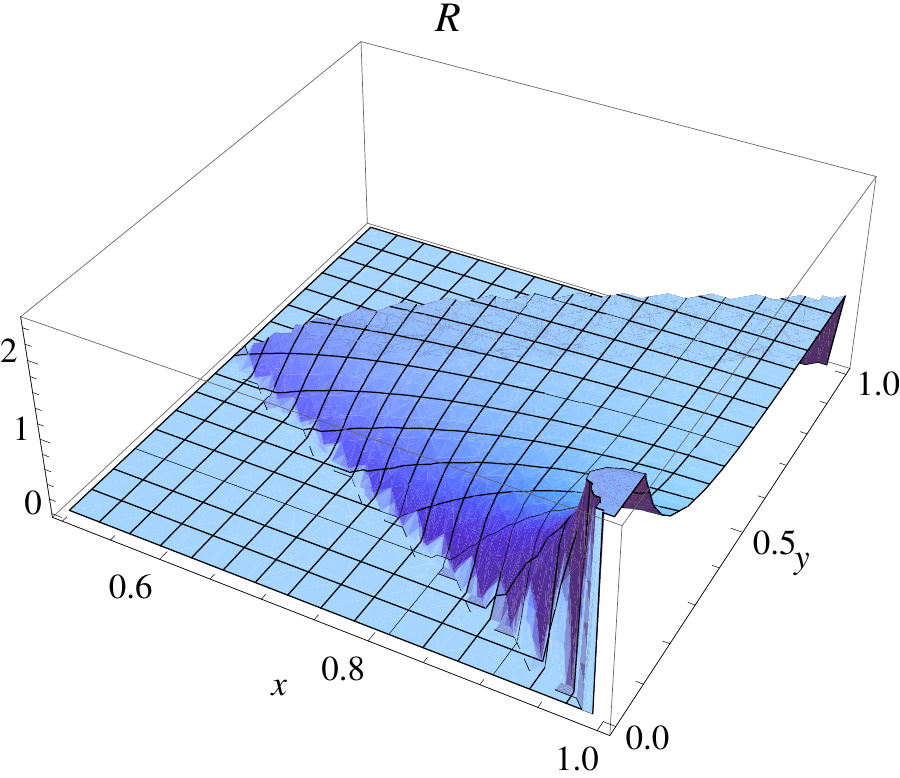}}\\
	\subfigure[]
	{\label{fig:5t+b}\includegraphics[width=55mm]{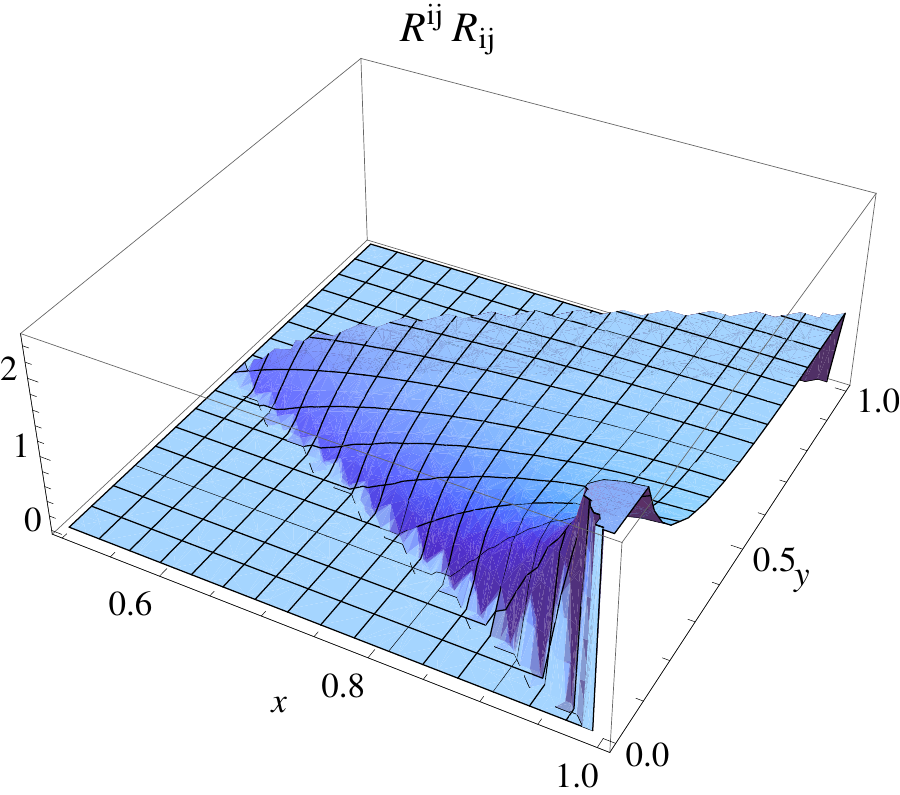}}\\
	\subfigure[]
	{\label{fig:5t+c}\includegraphics[width=55mm]{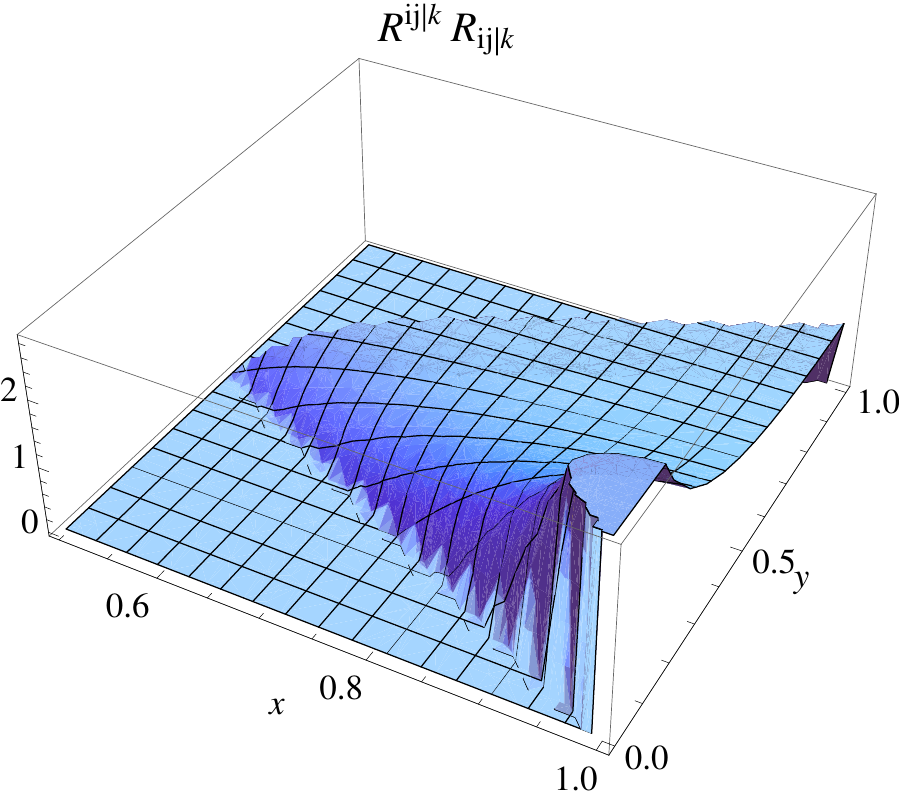}}\\
	\subfigure[]
	{\label{fig:5t+d}\includegraphics[width=55mm]{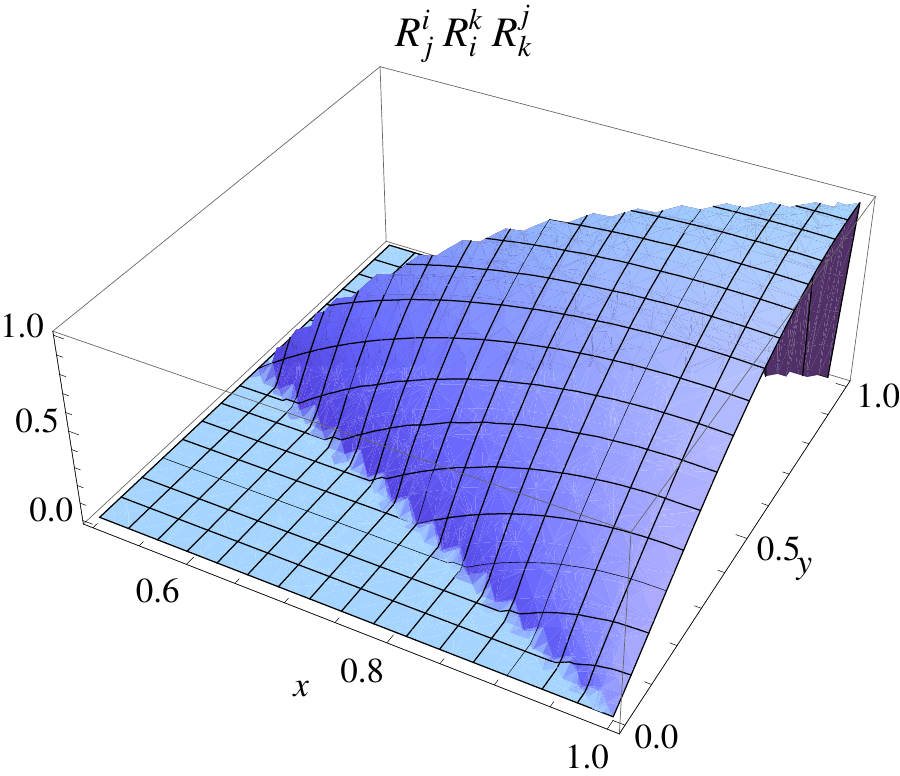}}
\caption{Shapes of $(k_1k_2k_3)^{-1} G_{+++} \lrp{\mathbf{K}}$ contributed by various terms. All are normalized to unity for equilateral limit. $x = k_2/k_1, y = k_3/k_1$.}
\lb{fig5t+}
\end{figure}
\begin{figure}
\centering
	\subfigure[]
	{\label{fig:5t-a}\includegraphics[width=55mm]{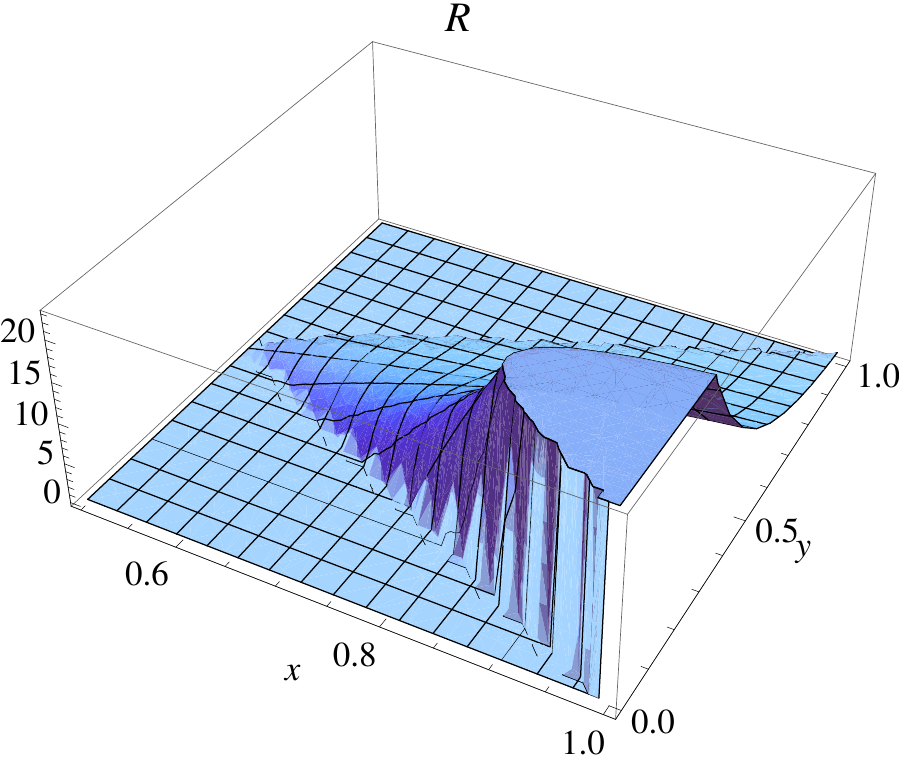}}\\
	\subfigure[]
	{\label{fig:5t-b}\includegraphics[width=55mm]{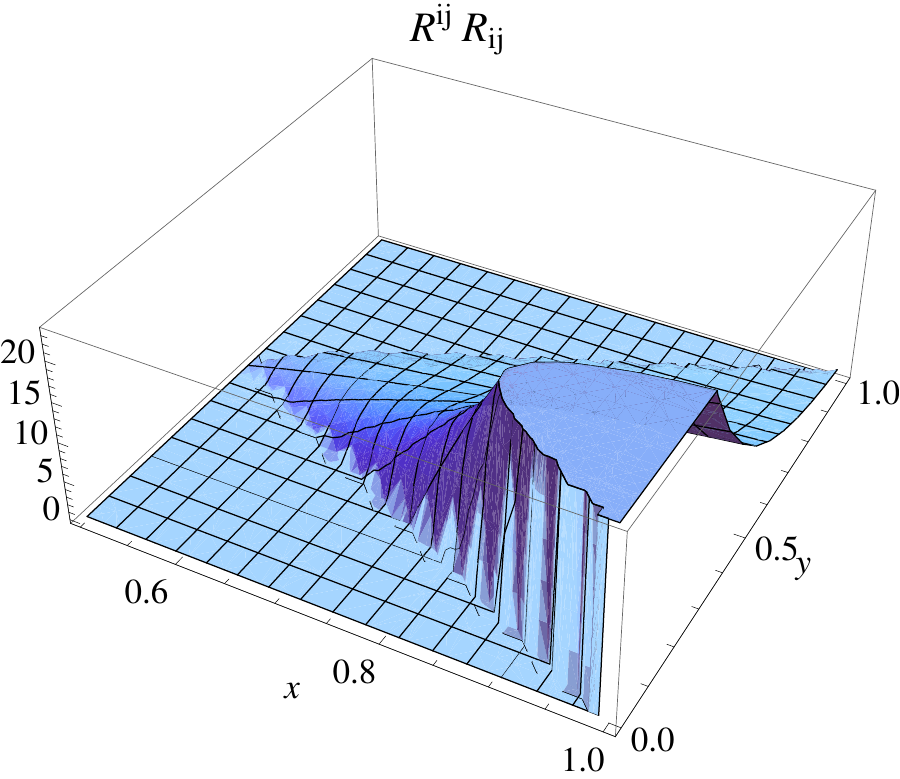}}\\
	\subfigure[]
	{\label{fig:5t-c}\includegraphics[width=55mm]{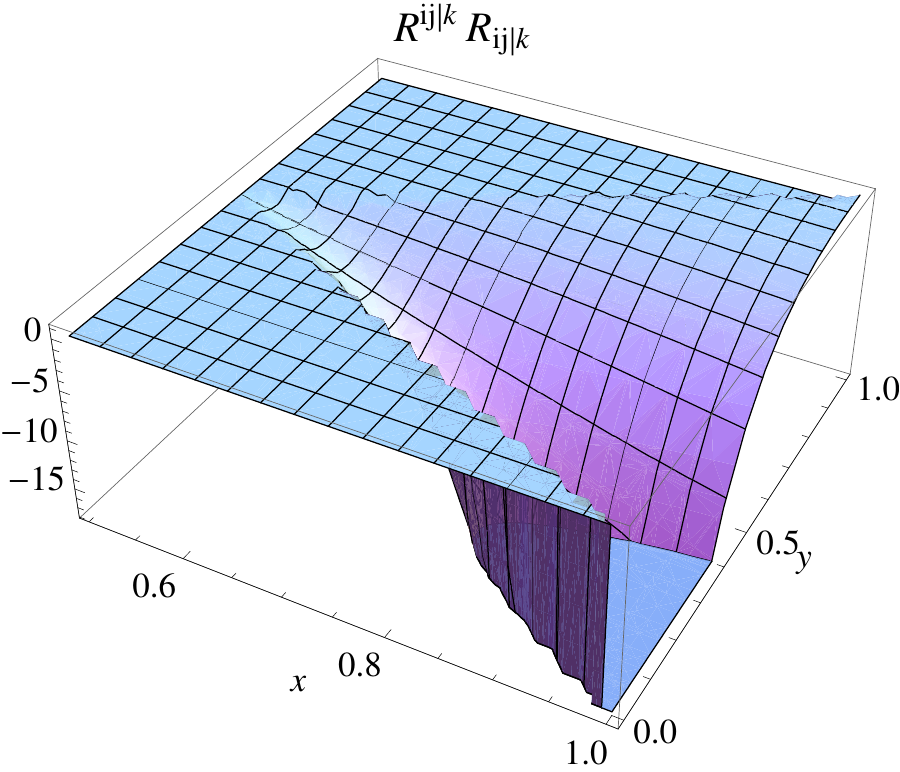}}\\
	\subfigure[]
	{\label{fig:5t-d}\includegraphics[width=55mm]{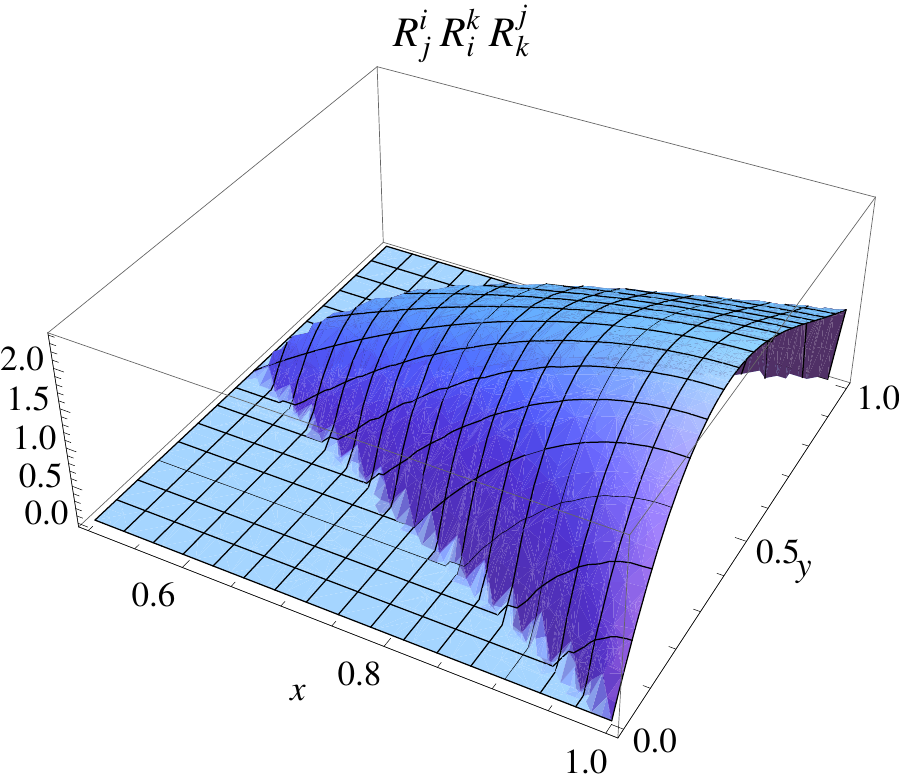}}
\caption{Shapes of $(k_1k_2k_3)^{-1} G_{++-} \lrp{\mathbf{K}}$ contributed by various terms. All are normalized to unity for equilateral limit. $x = k_2/k_1, y = k_3/k_1$.}
\lb{fig5t-}
\end{figure}
\begin{figure}
\centering
	\subfigure[Configuration 1]
	{\label{fig:5+basis.a}\includegraphics[width=55mm]{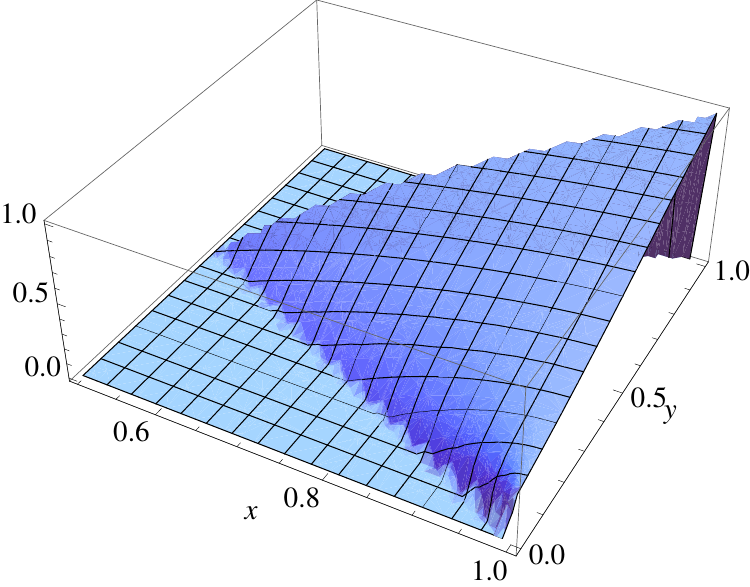}}\\
	\subfigure[Configuration 2]
	{\label{fig:5+basis.b}\includegraphics[width=55mm]{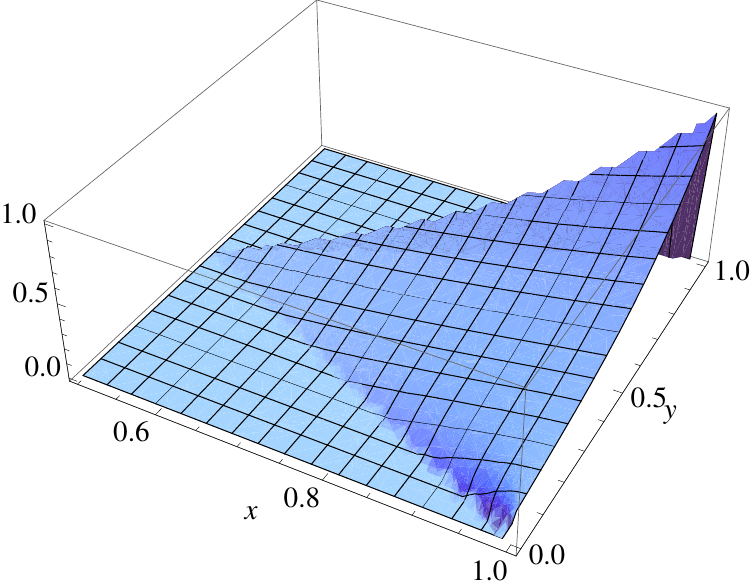}}\\
	\subfigure[Configuration 3]
	{\label{fig:5+basis.c}\includegraphics[width=55mm]{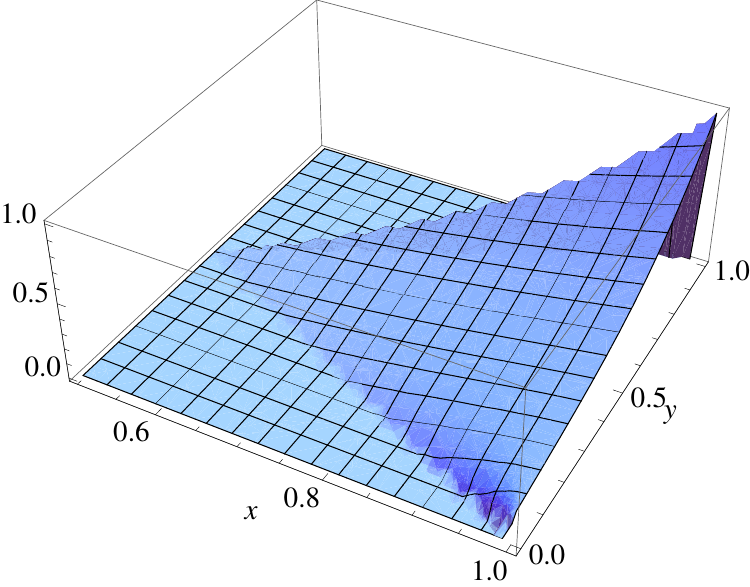}}\\
	\subfigure[Configuration 4]
	{\label{fig:5+basis.d}\includegraphics[width=55mm]{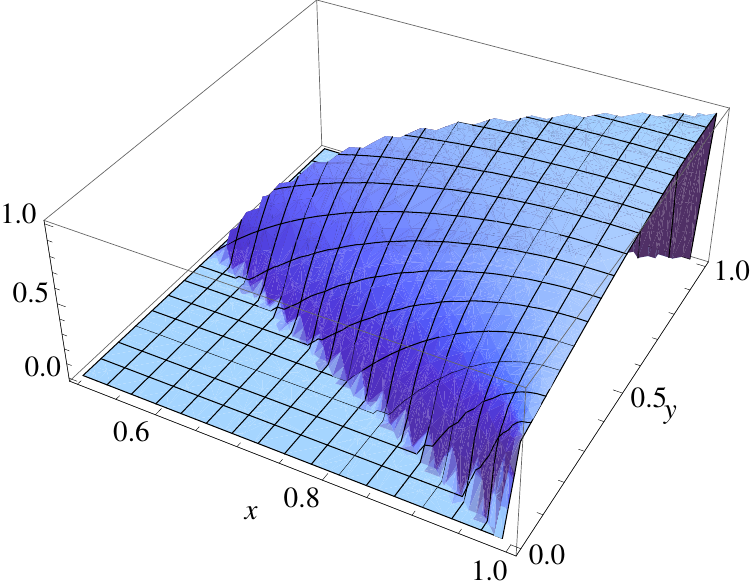}}
\caption{Shapes of the different configurations of the polarization tensors for $s_1 = s_2 =  s_3  = 1$. All are normalized to unity for equilateral limit. $x = k_2/k_1, y = k_3/k_1$.}
\lb{fig5t+.basis}
\end{figure}
\begin{figure}
\centering
	\subfigure[Configuration 1]
	{\label{fig:5-basis.a}\includegraphics[width=55mm]{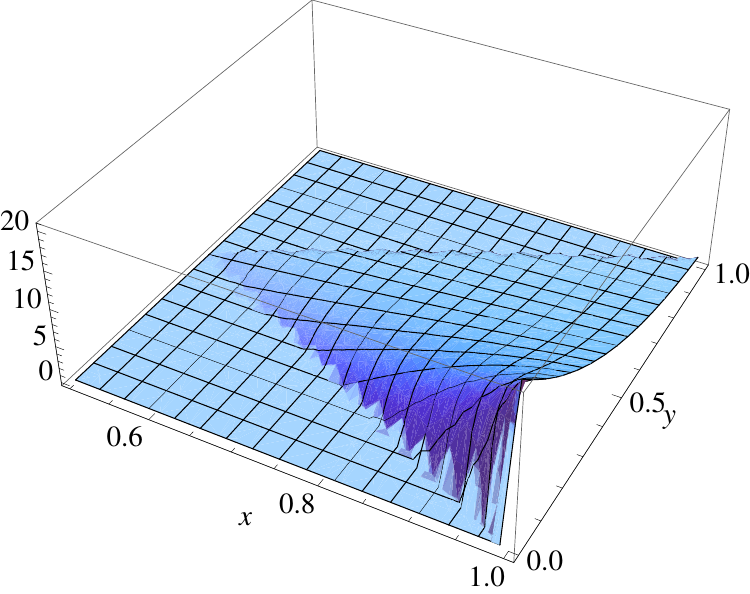}}\\
	\subfigure[Configuration 2]
	{\label{fig:5-basis.b}\includegraphics[width=55mm]{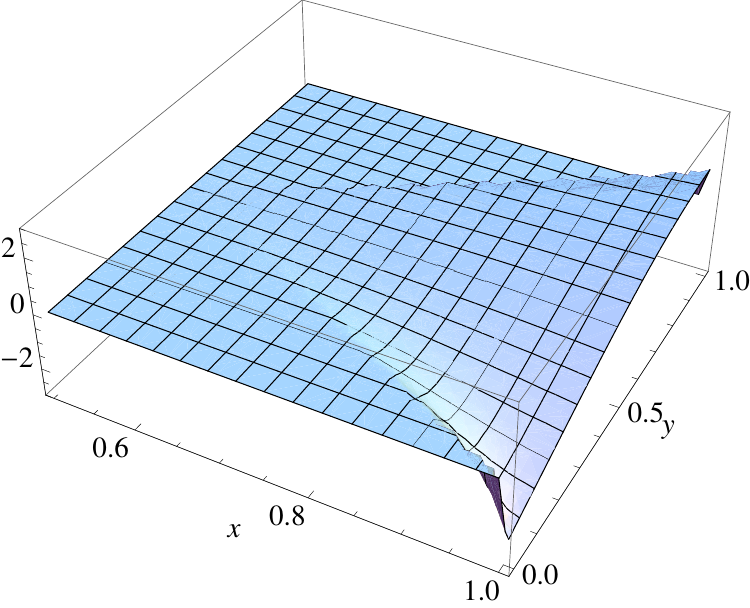}}\\
	\subfigure[Configuration 3]
	{\label{fig:5-basis.c}\includegraphics[width=55mm]{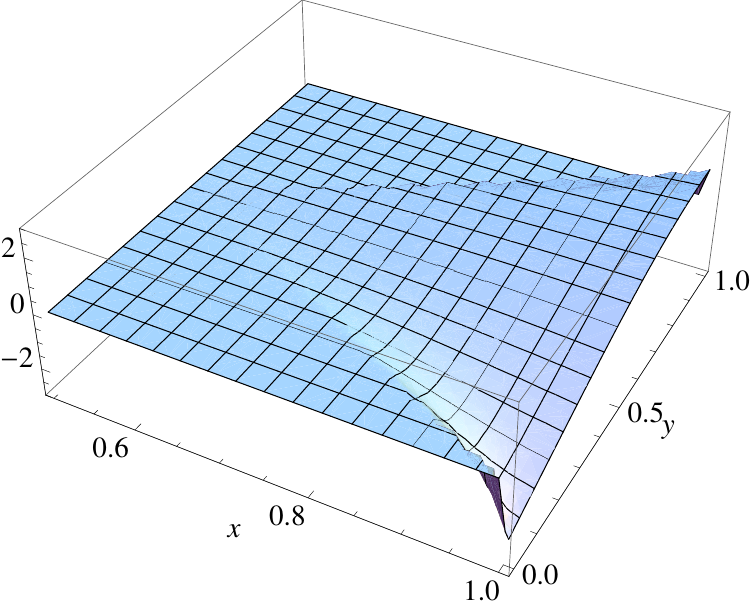}}\\
	\subfigure[Configuration 4]
	{\label{fig:5-basis.d}\includegraphics[width=55mm]{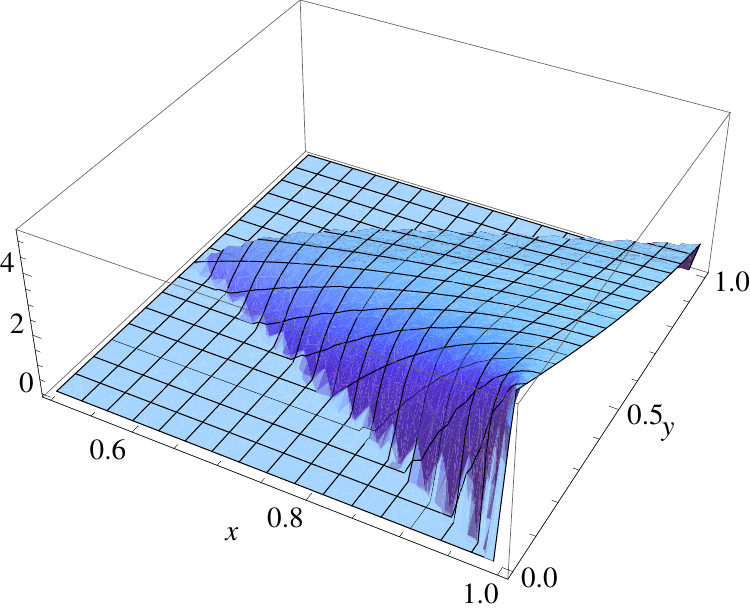}}
\caption{Shapes of the different configurations of the polarization tensors for $s_1 = s_2 =  -s_3  = 1$. All are normalized to unity for equilateral limit. $x = k_2/k_1, y = k_3/k_1$.}
\lb{fig5t-.basis}
\end{figure}
\section{Constraints from Planck Observations}
\renewcommand{\theequation}{6.\arabic{equation}} \setcounter{equation}{0}
Finally, we would like to comment on the recently-released results of Planck on primordial non-Gaussianity \cite{planck2013}.
As noted at the end of Section IV, the magnitude of the bispectrum is dependent on the mass scale $M_*$. Here we use the Planck result to obtain a constraint on $M_*$. 

Roughly speaking, the non-linearity parameter for bispectrum can be estimated as
\bq\lb{pl.1}
f^{\sst{\rm{T}}}_{\sst{\rm{NL}}} \sim \frac{\langle h h h\rangle}{\langle h h\rangle \langle h h\rangle} = \frac{\langle h h h\rangle}{\lrb{\Delta^2_{\rm{T}}}^2},
\eq
where $\Delta^2_{\rm T}$ is the scale-invariant power spectrum of the tensor perturbations.
  It is natural to assume that the bispectrum of the tensor perturbations has a lower magnitude than that of the scalar perturbations, that is,
\bq\lb{pl.2}
\langle h h h\rangle \lesssim \langle {\cal{R}}{\cal{R}}{\cal{R}}\rangle.
\eq
Thus,  we find that
\bqn\lb{pl.3-1}
&& \frac{\langle h h h\rangle}{\lrb{\Delta^2_{\rm{T}}}^2} \lesssim \frac{\langle {\cal{R}}{\cal{R}}{\cal{R}} \rangle}{\lrb{\Delta^2_{\rm{T}}}^2} 
= r^{-2} \frac{\langle {\cal{R}}{\cal{R}}{\cal{R}}\rangle}{\lrb{\Delta^2_{\cal{R}}}^2} = r^{-2} f^{\sst{\cal{R}}}_{\sst{\rm{NL}}}, \nb
\lb{pl.3-2}
\eqn
or
\bqn\lb{pl.3-1b}
  \langle h h h\rangle \lesssim r^{-2}{\lrb{\Delta^2_{\rm{T}}}^2}  f^{\sst{\cal{R}}}_{\sst{\rm{NL}}} =  \lrb{\Delta^2_{\cal{R}}}^2  f^{\sst{\cal{R}}}_{\sst{\rm{NL}}},
\eqn
the up bound of   which is  $\leq \lrp{4.9\times 10^{-5}}^2$ \cite{planck2013}, where $r= \Delta^2_{T}(k)/\Delta^2_{{\cal{R}}}(k)$.

The equilateral signal in our model takes the approximate value
\bq\lb{pl.4}
\langle h h h\rangle \sim \left| \frac{H C_{+}}{c_{\sst{\rm{T}}}}\right|^6 g_6,
\eq
where  we have taken $M_{\rm{pl}} \equiv 1$. The non-linearity parameter for the equilateral shape is constrained by the Planck result $f^{\sst{\cal{R}}}_{\sst{\rm{NL}}} = -42 \pm 75$, or $|f^{\sst{\cal{R}}}_{\sst{\rm{NL}}}| \lesssim 100$ \cite{planck2013}. This, together with Eqs.(\ref{pl.3-2}) and (\ref{pl.4}), requires
\bq\lb{pl.5}
\left| \frac{H C_{+}}{c_{\sst{\rm{T}}}}\right|^3 \sqrt{|g_6|} \lesssim 100 \lrp{4.9\times10^{-5}}^2 \lesssim 10^{-8}.
\eq

The normalization condition for the mode function $v_k$ in the relativistic region in Eq.(\ref{modes-r}) reads
\bq\lb{pl.6}
v_k v'^*_k - v^*_k v'_k = i \hbar,
\eq
which requires
\bq\lb{pl.6-1}
\frac{C_{+}}{\sqrt{c_{\sst{T}}}} \simeq 1.
\eq
%
Meantimes, a good IR behavior (so that the theory flows to general relativity  in this limit) requires that all the speeds of massless particles approach that of light, i.e.,   $c_{\sst{T}} \simeq 1$ \cite{Will}. Hence,  
the above equation implies
\bq
\frac{C_{+}}{c_{\sst{T}}} \simeq 1 \simeq \frac{C_{+}}{\sqrt{c_{\sst{T}}}}.
\eq

Apply this to condition (\ref{pl.5}), we have
\bq
H^3 \sqrt{|g_6|} \lesssim 10^{-8}.
\eq
Using the definition of $M_*$ in Eq. (\ref{m*}), this gives us the final constraint
\bq\lb{CM}
 \lrp{\frac{H}{M_*}}^2 \frac{H}{M_{\rm{pl}}} \lesssim 10^{-8}.
\eq
where  we have restored $M_{\rm{pl}}$ explicitly.

In the nonprojectable case without the extra U(1) symmetry, the frame effects imposed the most stringent constraint on the upper bound,   
$M_{*} \lesssim 10^{16}$ GeV \cite{BPS}, while in the case with the local U(1) symmetry, such effects have not been worked out yet, either with 
 or without the projectability condition. Assuming that this is also true in  the present case, the above condition implies that
$H/M_{*}  \lesssim 10^{-2}$. For $M_{*} \simeq M_{\rm pl}$, we find that $H/M_{*}  \lesssim 10^{-3}$. 

On the other hand,  in  \cite{Inflation} it was found that when $\epsilon_{HL} = (H/M_*)^2 \ll 1$, our model can reproduce 
the power spectrum given by the inflationary models in general relativity. 
Taking $\epsilon_{HL} \sim {\cal{O}} (10^{-2})$,  we find that constraint (\ref{CM}) gives $M_*/M_{\rm{pl}}\lesssim 10^{-5}$. 

\section{Conclusions and Remarks}

In this paper, we study 3-point correlation function of primordial gravitational waves generated during the de Sitter expansion of the universe
in the framework of the general covariant Ho\v{r}ava-Lifshitz gravity with the projectability condition and an arbitrary coupling constant $\lambda$.
We find that the interaction Hamiltonian, at cubic order ${\cal{O}}\left(h^3\right)$, receives contributions from the four dominant  terms: 
$$
R, \;\;\; R_{ij}R^{ij}, \;\;\; R^i_j R^j_k R^k_i, \;\;\; \left(\nabla^{i}R^{jk}\right)\left(\nabla_{i}R_{jk}\right).
$$
The Ricci scalar $R$ yields the same $k$-dependence as that in general relativity, i.e. its signal peaks at the squeezed limit 
regardless of the spins of the tensor fields, but with different magnitude due to coupling with the $U(1)$ field $A$ 
and a UV history when the dispersion relation is significantly different from the relativistic form. 
%
{Interestingly, the two terms $R_{ij}R^{ij}$ and $\left(\nabla^{i}R^{jk}\right)\left(\nabla_{i}R_{jk}\right)$ generate shapes similar to the $R$ term.
We show that this is due to the specific configuration of the polarization tensors [cf. configuration 1 in Eq. (\ref{confu})]. 

%
The term $R^i_j R^j_k R^k_i$ favors the equilateral shape when spins of the three tensor fields are the same but peaks in 
between the equilateral and squeezed limits when spins are mixed. Again we find that this is due to the effect of the polarization tensors: 
when spins are mixed, the product of the three polarization tensors strongly favors the squeezed shape.

{We also obtain a constraint on $M_*$ and $H$ [cf. Eq.(\ref{CM})] using the Planck  results, released recently \cite{planck2013}. 
This provides one of the strongest   constraints found so far in this version of the HL theory \cite{ZWWS,LW}. 

Finally, it should be noted that in performing the integration over the three mode functions to get the full expression of the bispectrum [cf. Eq.(\ref{bs})], 
we integrated over only the region when $k^2$ dominated the dispersion relation [cf. Eq.(\ref{modes-r})]. 
We gave a qualitative argument for the condition under which the UV history can be (partly) ignored, in leading order analysis. 
A quantitative study of the errors introduced with such ignorance will certainly deserve further analysis.

\section*{Acknowlodgements}

We would like to express our gratitude to Xian Gao and Satheeshkumar VH,  for valuable discussions and suggestions. 
AW and YQH thank Zhejiang University of Technology for hospitality.  AW is supported in part by the DOE  Grant, DE-FG02-10ER41692.


\end{document}